\documentclass{ws-procs9x6}
\usepackage{subfigure}

\begin{document}

\title{The search for neutrino-less double-beta decay}

\author{D. Tosi$^*$ on behalf of the EXO Collaboration}

\address{Physics Department, Stanford University,\\
Stanford, CA 94305, USA\\
$^*$E-mail: delia@slac.stanford.edu\\
}

\begin{abstract}
The search for neutrino-less double-beta decay has been a very active field of research in the last decade. Neutrino-less double-beta decay may answer essential open questions in neutrino physics. While double-beta decay accompanied by the emission of two neutrinos is allowed by the Standard Model, the neutrino-less process requires neutrinos to be Majorana particles. Detecting this decay could determine the nature of neutrinos, the neutrino effective mass, and the mass hierarchy. Xenon-136 and germanium-76 experiments currently have the best sensitivities. For both isotopes the lifetime of neutrino-less double-beta decay has been determined to be longer than $10^{25}$ years, several orders of magnitude slower than the slowest process that has been detected. In this paper the basic theory of neutrino-less double-beta decay and its relationship to the neutrino mass are discussed. A review of the experimental efforts underway to measure this decay is presented, along with the status and sensitivity of current and near future searches. 
\end{abstract}


\bodymatter
\section{Neutrino-less double-beta decay fundamentals}\label{sec1}

Neutrinos have been studied extensively since W. Pauli postulated them in 1930 in order to explain conservation of energy in beta decay. Despite the number of things we have learned about these elusive particles, they have still an unknown, and very small (but non-zero) mass. Not only, no right-handed neutrinos or left-handed neutrinos have been detected so far.  In 1937 Majorana hypothesized\cite{Majorana} the existence of self-conjugated fermions in parallel to self-conjugated bosons such as the Z boson and the photon. For a self-conjugated particle there is no distinction between a particle or its own anti-particle. Only neutrinos, being electrically neutral, are candidate members of this family of particles, which are now called Majorana fermions as opposed to Dirac fermions.
The idea of neutrinos being Majorana particles is attractive because it would allow, through the seesaw mechanism\cite{Valle(2006)}\,, for a justification of the small mass of these particles, which otherwise requires an unnatural coupling to the Higgs boson. It would also imply non conservation of the lepton number, and this could explain the asymmetry of baryons and anti-baryons.   

\begin{figure}[ht]
 \centering
 \subfigure[]{
  \includegraphics[scale=.292]{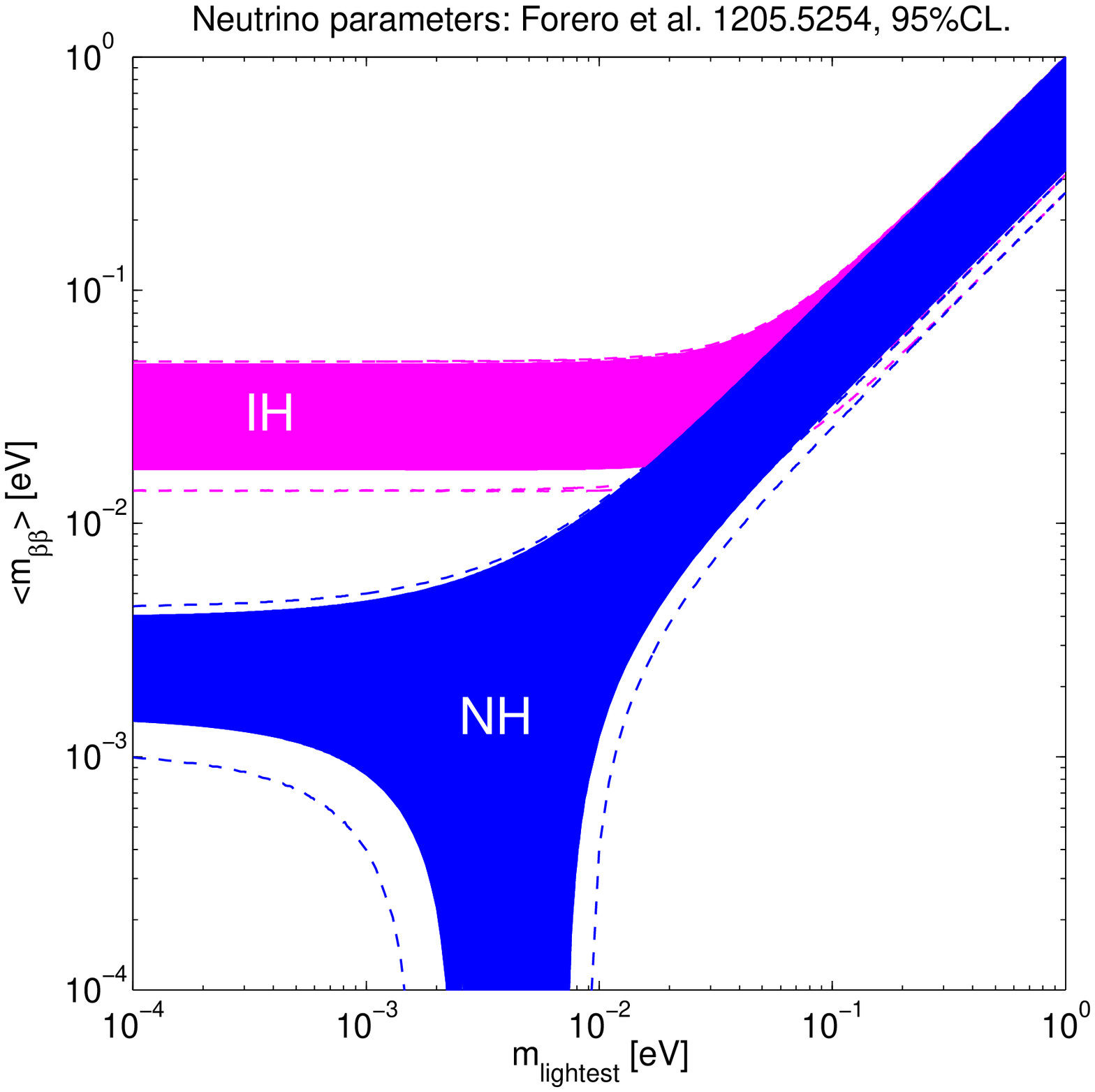}
   \label{fig:subfig1}
   }
 \subfigure[]{
  \includegraphics[scale=.292]{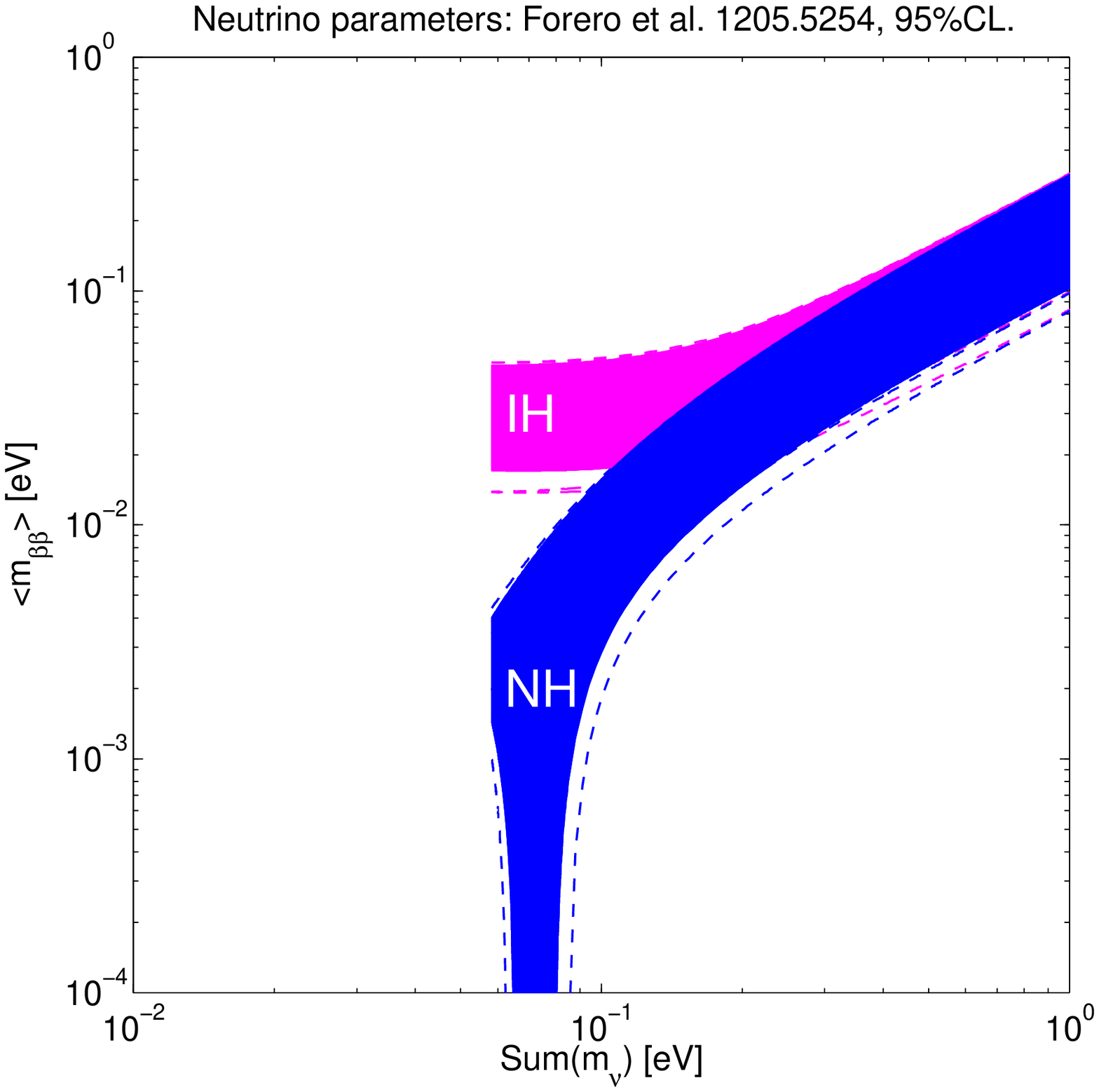}
   \label{fig:subfig2}
   }
 \label{fig:nuMass}
 \caption[Optional caption for list of figures]{%
  Neutrino effective mass for inverted or normal hierarchy with respect to the lightest mass \subref{fig:subfig1} and to the sum of neutrino masses \subref{fig:subfig2} according to up to date values of neutrino oscillations parameters \cite{Tortola:2012te}\,. The broken line shows the contours for the 95\% confidence level.}  
\end{figure}

The only experiments which can determine the Majorana/Dirac nature of neutrinos are those which search for neutrino-less double-beta decay.  
Double-beta decay ($2\nu\beta\beta$) is a radioactive standard second order process which involves two simultaneous beta decays, with the emission of two electron anti-neutrinos and two electrons.
The atom decaying undergoes the transformation X(A,Z)$\rightarrow$Y(A,Z+2). The process is observable only for a few isotopes where single beta decay is not allowed energetically. 
Even though the typical half lives for this process are very long (on the order of 10$^{18}$ to 10$^{21}$ years) this process has been measured in eleven isotopes. 
Neutrino-less double-beta decay ($0\nu\beta\beta$) is a variation of this decay which would happen only if neutrinos are Majorana particles. In this case,  the anti-neutrino produced by a beta decay could be absorbed by the other beta decay process as a neutrino. The signature of this event would be a peak in counts at the $2\nu\beta\beta$ decay energy spectrum upper end point, in a region of interest (ROI) defined around the maximum kinetic energy allowed for the emitted electrons (the endpoint or Q value). The width of the ROI depends on the energy resolution which is typical of the experiment. 
The expected rate of $0\nu\beta\beta$ is inversely proportional to the effective neutrino mass, which is defined as:
\begin{equation}
\langle m_{\beta\beta} \rangle= \vert\sum_i  U_{ei}^2m_{\nu i} e^{i\alpha_i}\vert
\label{eqn:effMass}
\end{equation}
where $U_{ei}$ are the elements of the first row of the Pontecorvo--Maki--Nakagawa--Sakata neutrino mixing matrix \cite{Maki:1962mu,Pontecorvo:1957qd} (containing the CP violation phase $\delta_{CP}$), and $\alpha_{i}$ are the Majorana phases (which are different from zero only if neutrinos are Majorana). 
Equation (\ref{eqn:effMass}) relates $\langle m_{\beta\beta} \rangle$ to the mass of the lightest neutrino eigenstate through the neutrino oscillation parameters, all of which are now measured, with the exception of $\delta_{CP}$. Figure \ref{fig:nuMass} represents \hbox{Eq. (\ref{eqn:effMass})} in the case of either inverted hierarchy (IH) or normal hierarchy (NH) depending on the unknown sign of $\Delta m_{23}$ and for all possible values of $\delta_{CP}$ and taking into account the uncertainty of the neutrino oscillation parameters \cite{Tortola:2012te}\,. 
%
%

The relation between 0$\nu\beta\beta$ half life and the neutrino effective mass (normalized to the electron mass $m_e$) is described by: 
\begin{equation}
\frac{1}{T^{0\nu\beta\beta}_{1/2}}= G^{0\nu}(Q,Z)\vert M^{0\nu}\vert^2 \left( {\frac{\langle m_{\beta\beta} \rangle}{m_e}} \right) ^2 
\label{eqn:t0nu}
\end{equation}
$G^{0\nu}(Q,Z)$ is the phase space factor and depends on the Q value and the atomic number Z and it is calculable \cite{Robertson(2013)}\,. $M^{0\nu}$ is the nuclear matrix element (NME) and its computation, quite complex, depends on the model used to approximate the wave functions of the two protons at the final state of the decay\cite{Vogel(2012)}\,. The state of art for this calculation is reviewed in Ref.~\refcite{Barea(2013)}. Currently, NME values are estimated within a factor of about two.

\section{A selection of experiments}\label{sec2}
In the past decade, many experiments have tried to measure the half life of double-beta decay. 
Historically, there have been two types of experiments:
\begin{itemlist} 
\item{One method is based on the identification of atoms in the final state (A,Z+2) in a mass of element (A,Z). This method does not require waiting any time comparable to the half life but does not distinguish which type of process ($0\nu\beta\beta$ or $2\nu\beta\beta$) has transformed those atoms in the final state.}
\item{The other method relies on observing a large mass of a particular isotope and detect the decays while happening, measuring the energy of the electrons emitted during the decay. Once a spectrum with sufficiently high statistics and sufficiently low background has been collected, an excess of events can be searched for at the endpoint. To achieve low background, these experiments need to be done underground (to reduce cosmic rays) with additional shielding (to reduce radiation from rocks).} 
\end{itemlist}
Currently no method combining the two techniques has been developed. If this was achieved, the background in the ROI would be reduced to $2\nu\beta\beta$ decay only. The EXO Collaboration is engaged in R\&D to tag the final state of an atom after observing a decay in xenon-136. Intermediate techniques for background reduction are already in operation and a few experiments are able to distinguish the topology (single-site vs. multi-site) of the event and identify the decay type ($\alpha,\beta,\gamma$) by tracking or imaging the charge created by the decay event. 

Another essential feature of double-beta decay experiments is the isotope choice. A claim of the discovery of $0\nu\beta\beta$ decay will need to be proved in more than one isotope. 
The ideal isotope for a $0\nu\beta\beta$ search would have the following features:
\begin{itemize}
\item{high endpoint, above 2.5 MeV}
\item{short half life}
\item{high natural abundance}
\item{large ratio of $0\nu$ to $2\nu$ events}
\item{easy/inexpensive to enrich and to obtain in large mass}
\item{large ratio of $0\nu\beta\beta$/$2\nu\beta\beta$ events in the ROI} 
\end{itemize}
Unfortunately none of the eleven isotopes for which the search is in principle possible has all these features. For an overview see Ref.~\refcite{Robertson(2013), Cremonesi(2013),Biller(2013)}.
Among the candidate isotopes, germanium-76, xenon-136, tellurium-130 are used in several experiments -- these will be reviewed here. Investigation has begun more recently of neodymium-150, cadmium-116, selenium-82,  molybdenum-100, calcium-48.

Xenon-136 has a natural abundance of 8.9\% but can be enriched relatively easily; it can also be purified continuously and reused. The endpoint (\hbox{Q = 2458 keV}) is higher than most naturally occurring backgrounds. The cosmogenic activation is minimal, since there are no long-lived radioactive isotopes; moreover, thanks to its high atomic number (\hbox{Z = 54}) and its high density (\hbox{$\sim$ 3~g/cm$^3$}), liquid xenon provides self shielding from background $\gamma$ rays. 
A valuable property of $^{136}$Xe is that the scintillation light and the ionization charge produced by a decay event are anti-correlated, and if both are measured the energy resolution improves considerably. Finally, the background can potentially be reduced by tagging the Ba$^{++}$ daughter of the decay.
Currently, three experiments are using $^{136}$Xe to search for $0\nu\beta\beta$ decay: EXO-200, NEXT-100 and KamLAND-Zen. EXO-200 is located at the Waste Isolation Pilot Plant near Carlsbad, NM. The experiment is fully described in Ref.~\refcite{EXO200_det1}. It uses a time projection chamber (TPC)\cite{TPC}\,, which contains 175~kg of liquid xenon (LXe), enriched to 80.672$\pm$0.14$\%$ in $^{136}$Xe. The TPC, a 1.4~mm thick cylindrical copper vessel of 40~cm diameter and 44~cm length, is divided in two halves by the cathode which is biased at -8~kV. Each side of the TPC is equipped with two sets of wires and a few hundreds large area avalanche photo-diodes (LAAPDs)\cite{APDs}\,. When a radioactive decay deposits energy in the LXe, charge and scintillation light are created. A set of wires detects the induction signal generated by the charge drifting in the electric field; the other set collects the charge. The LAAPDs detect the light. Using all three signals the event location, topology and energy are reconstructed. The energy resolution achieved is 1.84\% ($\sim$4.3\% FWHM) at the Q~value\cite{EXO200_2nu_2013}\,. The ability to reconstruct topology helps distinguishing background $\gamma$ events (which are mostly multi-site) from beta events (which are mostly single-site).  Xenon is continuously recirculated and purified in gas phase, in order to maintain as low impurities as possible and to collect all the charge of each event. Since the start of data taking in 2011, the experiment has published the first measurement of $2\nu\beta\beta$ in $^{136}$Xe \cite{EXO200_2nu_2011}\,, a limit on the half life of $0\nu\beta\beta$ \cite{EXO_0nu} (see Table \ref{tab:limits}), and more recently, the most precise measurement of the $2\nu\beta\beta$ rate for any isotope \cite{EXO200_2nu_2013}\,. The next generation detector, nEXO, is currently being designed. This will be similar to EXO-200, and will feature a copper TPC filled with LXe submerged in HFE\cite{HFE}\,. The total mass of Xe will be about 5 tonnes, enriched at 90\% in $^{136}$Xe. The detector will be designed to keep open the possibility of adding instrumentation to tag barium atoms at a later stage. The expected sensitivity is about \hbox{3 $\cdot$ 10$^{27}$ yr} after five years and will go beyond \hbox{2 $\cdot$ 10$^{28}$ yr} if barium tagging is implemented. This will probe the inverted hierarchy. 

Another experiment based on xenon is NEXT-100, expected to start taking data in 2014-2015. NEXT-100\cite{NEXT(2013)}\,, located in the Laboratorio Subterraneo de Canfranc, uses a gas TCP filled with 100~kg of enriched Xe at \hbox{15 bar} pressure. The TPC is equipped with photo-multiplier-tubes (PMTs) on one side and with an array of multi-pixel-photon-counter (MPPCs) on the other side. An event creates prompt scintillation light and ionization. The charge is drifted to a higher electric field (in the proximity of the MPPCs) where it produces another pulse of light by electroluminescence. The PMTs detect both scintillation (used to mark the start of the event) and electroluminescence (used for energy reconstruction), while the MPPCs provides tracking and topology reconstruction through detection of the electroluminescence signal. Using prototypes the NEXT collaboration has demonstrated the ability both to achieve very good energy resolution \cite{NEXT_DEMO(2012),NEXT_DEMO(2013)} (extrapolated to 0.5\%-0.7\% FWHM at the endpoint) and to reconstruct tracks at different energies. For more information regarding NEXT, see Ref.~\refcite{NEXTproc}.

KamLAND-Zen was built as a modification of the low energy anti-neutrino detector KamLAND, located in the Kamioka mine at Hiba City in Japan. 300~kg of $^{136}$Xe are diluted in Liquid Scintillator (LS) for a total mass of 13~ton. This material is contained into a 1.54~m radius spherical inner balloon (IB) which is suspended inside a 6.5~m radius spherical outer balloon (OB) containing 1~kton LS. 
KamLAND-Zen published a measurement for the 2$\nu\beta\beta$ half life in $^{136}$Xe \cite{KLZ_2nu} and a limit for the $0\nu\beta\beta$ half life which is currently the best limit for $^{136}$Xe \cite{KLZ_0nu} (see Table \ref{tab:limits}). The largest backgrounds measured in the ROI are $^{110m}$Ag and $^{214}$Bi from the U-decay chain. The latter seems to concentrate on the IB welding lines, together with other contaminants (responsible for events at lower energy) likely coming from Fukushima fallout.  $^{110m}$Ag may also come from the fallout or from cosmogenic production by Xe spallation. Purification of the Xe and of the liquid scintillator is underway and a new run is foreseen to start in November 2013. An upgrade of the detector will follow and will include the installation of a new internal balloon with 800~kg of $^{136}$Xe. This is planned to happen around 2016 and  will provide the necessary sensitivity to probe the inverted hierarchy. 

Germanium-76 has been used for decades in the search of $0\nu\beta\beta$ decay.  In addition to offering the advantage of a very good energy resolution (as low as 0.2\% FWHM at the Q value of 2039~keV), germanium-76 has attracted interest due to a controversial claim for evidence of $0\nu\beta\beta$ decay \cite{Klapdor} based on the data from the Heidelberg-Moscow collaboration. 
The Heidelberg-Moscow collaboration reported the limit T$_{1/2}$ \textgreater 1.9 $\cdot$ 10$^{25}$ yr in 2001\cite{HM_Coll}\,. In the same year, Klapdor-Kleingrothaus {\it et al.} claimed the observation of $0\nu\beta\beta$ decay of $^{76}$Ge with a half life of 1.5 $\cdot$ 10$^{25}$ yr \cite{Klapdor_2001}\,, changed later in 2004 to \hbox{1.19 $\cdot$ 10$^{25}$ yr} \cite{Klapdor_2004}\,,  and again in 2006 to \hbox{2.23$^{+0.44}_{-0.31}$ $\cdot$ 10$^{25}$ yr} \cite{Klapdor_2006}\,. All of this was in disagreement with the rest of the collaboration \cite{HM_2005}\,.

Currently two experiments are based on $^{76}$Ge: the Majorana demonstrator and GERDA. GERDA \cite{GERDA}, located at the Gran Sasso Laboratory, uses bare Ge detectors enriched to 86\% in $^{76}$Ge submerged in liquid argon (LAr), which serves both as cooling agent and shield (along with a shell of high purity water). Its first round of data (phase I) was taken between November 2011 and May 2013, with a mass of 18~kg distributed over two different types of detectors (semi-coaxial high purity germanium or HPGe and broad energy germanium or BEGe), for a total exposure was \hbox{21.6~kg$\cdot$yr}. No events have been observed, setting the current best limit for T$^{0\nu\beta\beta}_{1/2}$ in $^{76}$Ge (see Table \ref{tab:limits}). A background level of \hbox{$1\cdot10^{-2}$cnts/(keV$\cdot$kg$\cdot$yr)} has been achieved\cite{GERDA_0nu}\,. GERDA is going forward with its planned phase II, which promises a background lower by a factor 10, an improved energy resolution (thanks to better detectors) and a sensitivity of \hbox{$10^{26}$ yr} (thanks to an additional 20~kg  of germanium). For more details about GERDA see also Ref.~\refcite{GERDAproc}.
 
The Majorana demonstrator\cite{MajoranaDemonstrator}\,, located at the Sanford Underground Research Facility, features 30~kg of germanium enriched to 86\% in $^{76}$Ge and 10~kg of natural Ge. The Ge diodes (located in two independent vacuum cryostats) have low energy threshold and very good energy resolution, and can also be used for dark matter search. The goal of the experiment is to achieve a background as low as 3~cnts/tonne/year in a ROI of width 4~keV using a shield made of lead, oxygen-free copper, electro-formed copper, and scintillator paddles. The experiment construction should be completed by late 2014. 
GERDA and Majorana collaborations are working together to prepare for a ton scale detector.  


\begin{figure}[htbp]
\begin{center}
\psfig{file=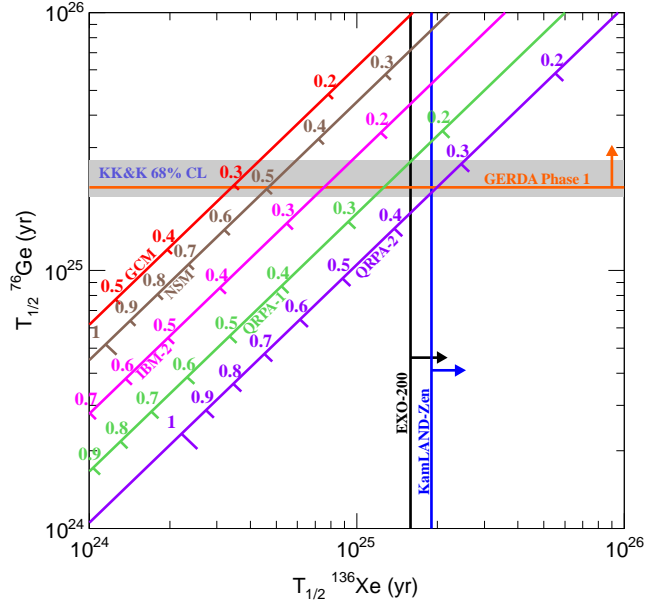,width=3.5in}
\caption{Relation between the $T^{0\nu\beta\beta}_{1/2}$ in $^{76}$Ge and in $^{136}$Xe for different nuclear matrix element models (GCM\cite{GCM}\,, NSM \cite{NSM}\,, IBM-2 \cite{IBM-2}\,, RQRPA-1 \cite{QRPA1} and QRPA-2 \cite{QRPA2}). For each nuclear matrix element, the values of $<m_{\beta\beta}>$ (eV) are indicated on the corresponding line.  The claimed detection \cite{Klapdor} is represented by the grey band and compared with the best limit $^{76}$Ge \cite{GERDA_0nu} and the best limits in $^{136}$Xe \cite{EXO_0nu,KLZ_0nu}\,.}
\label{fig:GeVsXe}
\end{center}
\end{figure}


The recent results in Xe and Ge can be compared and displayed on the same graph for different nuclear matrix elements. As shown in Fig. \ref{fig:GeVsXe} the claim \cite{Klapdor} is excluded for most matrix elements.

The use of tellurium-130 in the search for $0\nu\beta\beta$ decay has been explored since the mid nineties by the CUORE collaboration, in the Laboratories of Gran Sasso. Tellurium-130 offers the advantage of a natural abundance of  34.2\%. 
It's often used in the stable compound TeO$_2$ which has good mechanical and thermal properties. The crystals are used as bolometers, cooled down to cryogenic temperatures. When a decay occurs inside the crystal, it creates phonons. This causes a temperature increase detectable by thermistors. CUORICINO \cite{CUORICINO}, a demonstrator for CUORE made of 62 TeO$_2$ crystals with a total mass of 41~kg, operated between 2003 and 2009 and established what currently is the best limit for T$^{0\nu\beta\beta}$ in this isotope (see Table \ref{tab:limits}). CUORE-0, one CUORE-like tower with 13 planes of 4 crystals each, has just started operating. It has a similar mass to CUORICINO but thanks to intensive surface cleaning procedures has achieved a background of  0.074 $\pm$ 0.012 cnts/keV/kg/yr in the ROI --- a factor of six better than CUORICINO's. CUORE, with a mass of about 200~kg of $^{130}$Te, is expected to start taking data in 2015 and reach a sensitivity of \hbox{1.6$\cdot$10$^{25}$yr} after running five years \cite{CUORE}\,.

SNO+, located in Sudbury, Canada, has joined the search for $0\nu\beta\beta$ decay after replacing the heavy water of the SNO experiment with liquid scintillator. Recently the collaboration has switched from $^{150}$Nd to $^{130}$Te as isotope to add to its scintillator. Construction is ongoing and data taking should start in 2014. More about the status of this experiment can be found in these proceedings at Ref.~\refcite{SNO+proceedings}. 

A summary of the current limits and future sensitivities are presented in Tables \ref{tab:limits}--\ref{tab:sensitivities}.

\begin{table}
\tbl{Summary of T$^{0\nu\beta\beta}_{1/2}$ lower limits}
{\begin{tabular}{@{}cccc@{}}\toprule
Isotope & Experiment &T$^{0\nu\beta\beta}_{1/2} $[yr]   & $\langle m_{\beta\beta} \rangle$ [meV] \\
\ \
\colrule
$^{136}$Xe & EXO-200 & \textgreater 1.6 $\cdot$ 10$^{25}$ & \textless 140--380  \\
$^{136}$Xe & KamLAND-Zen & \textgreater 1.9 $\cdot$ 10$^{25}$ & \textless 120--250 \\
$^{76}$Ge & GERDA phase I  &\textgreater 2.1 $\cdot$ 10$^{25}$ & \textless 200--400 \\
$^{130}$Te & CUORICINO  &\textgreater 2.8 $\cdot$ 10$^{24}$ & \textless 300--700 \\
\botrule
\end{tabular}
}
\label{tab:limits}
\end{table}

\begin{table}
\tbl{Sensitivity of future experiments}
{\begin{tabular}{@{}cccc@{}}\toprule
Isotope & Experiment &T$^{0\nu\beta\beta}_{1/2} $  & $\langle m_{\beta\beta} \rangle$ \\
 &  &sensitivity [yr]   &  sensitivity [meV] \\
\ \
\colrule
$^{136}$Xe & EXO-200 (4 yr) & 5.5$\cdot$10$^{25}$ & 75--200  \\
$^{136}$Xe & nEXO (5 yr) & 3$\cdot$10$^{27}$ & 12--29  \\
$^{136}$Xe & nEXO (5 yr + 5 yr w/ Ba tagging) & 2.1 $\cdot$10$^{28}$ & 5--11  \\
$^{136}$Xe & KamLAND-Zen (300~kg, 3 yr)& 2$\cdot$10$^{26}$ &45-110 \\
$^{136}$Xe & KamLAND2-Zen (1~ton, post 2016)&  IH & IH  \\
$^{76}$Ge & GERDA  phase II  & 2$\cdot$10$^{26}$ &90--290 \\
$^{130}$Te & CUORE-0 (2 yr) &  5.9$\cdot$10$^{24}$ &204--533 \\
$^{130}$Te & CUORE (5 yr) & 9.5$\cdot$10$^{25}$ & 51--133 \\
$^{130}$Te & SNO+ & 4$\cdot$10$^{25}$ & 70--140 \\
\botrule
\end{tabular}
}
\label{tab:sensitivities}
\end{table}

\section{Conclusions}\label{sec3}
Neutrino-less double-beta decay has been a very active field of physics in the past decade, and promises to be even more active in the next decade. The current generation experiments (EXO-200, KamLAND-Zen, GERDA phase\,I) have nearly disproved the only claim for a signal and will keep taking data and improve their limits. In the next decade the mass sensitivity of near future experiments will probe a critical region in the parameter space of effective Majorana mass vs. lightest neutrino mass.  If the hierarchy is inverted, it will be possible to make a discovery, or exclude the Majorana nature of neutrinos. If the hierarchy is normal, the discovery potential of neutrino-less double-beta decay experiments will depend on the lightest neutrino mass.    

\section*{Acknowledgments}\label{sec4}
The author wishes to thank the conference organizers for the opportunity of preparing this review and J.~Detwiler, J.~J.~Gomez-Cadenas, E.~Previtali, R.~Maruyama, N.~Tolich, for providing information regarding the experiments.

\bibliographystyle{ws-procs9x6}

\end{document}